\documentclass[a4paper,11pt,dvips]{article}
\usepackage{graphicx}
\usepackage{amsmath}
\usepackage{amssymb}
\usepackage{latexsym}
\usepackage{subcaption}
\usepackage[colorlinks]{hyperref}
\usepackage{color}
\usepackage{enumerate}
\usepackage{enumitem}
\usepackage{cite}
\usepackage{makeidx}
\usepackage{colortbl}

\newcommand{\bra}{\begin{array}}
\newcommand{\era}{\end{array}}
\newcommand{\beq}{\begin{equation}}
\newcommand{\eeq}{\end{equation}}
\newcommand{\bqr}{\begin{eqnarray}}
\newcommand{\eqr}{\end{eqnarray}}

\def\BC{\bb C}
\def\_\BC{\bbi C}

\def\no2 {{\textstyle{n\over 2}}}

\newcommand{\lb}{\label}

\begin{document}
\begin{titlepage}
\setcounter{page}{1}
\renewcommand{\thefootnote}{\fnsymbol{footnote}}

\begin{flushright}
\end{flushright}

\vspace{5mm}
\begin{center}

{\Large \bf {Goos-H\"anchen Shifts in Graphene-Based Linear
 Barrier}}

\vspace{5mm} {\bf Miloud Mekkaoui}$^{a}$, {\bf Radouane El Kinani
}$^{a}$ and {\bf Ahmed Jellal\footnote{\sf a.jellal@ucd.ac.ma}}$^{a,b}$

\vspace{5mm}

{$^{a}$\em Laboratory of Theoretical Physics,  
Faculty of Sciences, Choua\"ib Doukkali University},\\
{\em PO Box 20, 24000 El Jadida, Morocco}

{$^b$\em Saudi Center for Theoretical Physics, Dhahran, Saudi Arabia}


\vspace{3cm}

\begin{abstract}

Using the energy spectrum of a system
made of graphene
subjected to a linear barrier potential, we study
the Goos-H\"anshen shifts. The calculation is done
by first determining  the corresponding
phase shifts via the transmission and reflection probabilities. Numerical analysis
shows that the Goos-H\"anshen shifts depend strongly on
the incident energy,
barrier height and width, and vary positively or negatively
under suitable conditions.

\end{abstract}
\end{center}

\vspace{3cm}


\end{titlepage}


\section{Introduction}


Graphene is a honeycomb lattice in two dimensions, experimentally obtained in
2014 by Manchester group~\cite{Novoselov}. From time being graphene gives a good
laboratory to test many theories
like for instance quantum electrodynamics where Klein paradox
takes place. This can experimentally be realized in graphene systems
by considering Dirac fermions scattered by barrier potentials~\cite{Stander}.
There is a big progress in studying quantum phenomena  in graphene systems among them
we cite the quantum version of the
Goos-H\"anchen
(GH) effect originating from the reflection of particles from
interfaces \cite{Goos}. Many works in various graphene-based nanostructures,
including single \cite{Chen15}, double barrier \cite{Song16}, and
superlattices \cite{Chen18}, showed that the GH
shifts can be enhanced by the transmission resonances and
controlled by varying the electrostatic potential and induced gap
\cite{Chen15}. Similar to those in semiconductors, the GH shifts
in graphene can also be modulated by electric and magnetic
barriers \cite{Sharma19}. 
It has
been reported that the GH shift plays an important role in the
group velocity of quasiparticles along interfaces of graphene p-n
junctions \cite{Beenakker,Zhao11}.

In our previous work \cite{HBahlouli}, we have
 solved the 2D Dirac equation describing graphene in the presence of a linear vector potential. The
discretization of the transverse momentum due to the infinite mass boundary condition reduced our 2D
Dirac equation to an effective massive 1D Dirac equation with an effective mass equal to the quantized
transverse momentum. We have used both a numerical Poincaré map approach, based on space discretization of
the original Dirac equation, and a direct analytical method. These two approaches have been used to study
tunneling phenomena through a biased graphene strip. We have showed that
the numerical results generated by the Poincaré
map are in complete agreement with the analytical results.
In the second one \cite{Wang},
we have explored the zero, positive and negative quantum  GH shifts of the transmitted
Dirac carriers in graphene through a potential barrier with vertical magnetic field. Numerical results
show that only one energy position at the zero GH shift exists and is highly dependent on the
$y$-directional wave vector, the energy gap, the magnetic field and the potential.
We have showed that the positive and negative GH
shifts happen when the incident energy is more and less than the energy position at the zero GH shift,
respectively.
In addition, we found that there are two values of potential at the zero GH shifts, where a
potential window can always keep the positive GH shifts.

Motivated by  our previous work \cite{HBahlouli,Wang},
we consider Dirac fermions in graphene subjected to a linear barrier
potential and study the GH shifts.
From the solution of the energy spectrum we show how to derive
the GH shifts as function of different physical parameters
based on the phase shifts in transmission and reflection.
To give a better
understanding of our results, we give a numerical study based on
different choices of the physical parameters. Among the obtained results
we show that GH shifts can be controlled by linear barrier potential.

The present paper is organized as follows. In section 2,
we set our problem and write down the corresponding Hamiltonian
as well as the solutions of the energy spectrum for different regions
composing our system.
These will be used in section 3 to determine the transmission and
reflection probabilities from which we derive the phase shifts.
Using standard definition, we end up with the GH shifts in terms of
the physical parameters characterizing our system.
In
section 4, we numerically analyze and discuss the phase shifts, GH shifts and transmission
by considering suitable choices of the physical parameters.
We
conclude our results in the final section.

\section{ Energy spectrum of the system}

We consider
massless Dirac fermions
through a graphene scattered by linear barrier potential, as shown in Figure \ref{db.1}a, with
incident energy $E$ and angle $\phi_1$ with
respect to the $x$-direction, while they are free in the $y$-direction.
Our system is made of three regions denoted by $j$ = $1, 2, 3$ and
each region is characterized by a given potential.
The barrier regions are formally described by
the Dirac-like Hamiltonian
\begin{equation}\lb{Ham1}
H=v_{F} {\boldsymbol{\sigma}}\cdot\textbf{p}+V(x){\mathbb I}_{2}
\end{equation}
where ${v_{F}\approx 10^6 m/s}$  is the Fermi velocity,
${{\boldsymbol{\sigma}}=(\sigma_{x},\sigma_{y})}$ are the Pauli
matrices, $\textbf{p}=-i\hbar(\partial_{x},
\partial_{y})$, ${\mathbb I}_{2}$ the $2 \times 2$ unit matrix,
the electrostatic potential $V(x)=V_{j}$ in each scattering
region is given by
\begin{equation}
V(x)=V_{j}=
\left\{%
\begin{array}{ll}
    -\Lambda x+V_0, & \hbox{$0\leq x\leq d$} \\
    0, & \hbox{otherwise} \\
\end{array}%
\right.
\end{equation}
with $\Lambda=\frac{V_0}{d}$,  as presented schematically
in Figure \ref{db.1}a. We illustrate the negative and positive GH shifts of
Dirac fermions in transmission through the graphene linear
barrier in Figure \ref{db.1}b.

\begin{figure}[!ht]
\centering
\includegraphics[width=8cm, height=5.5cm]{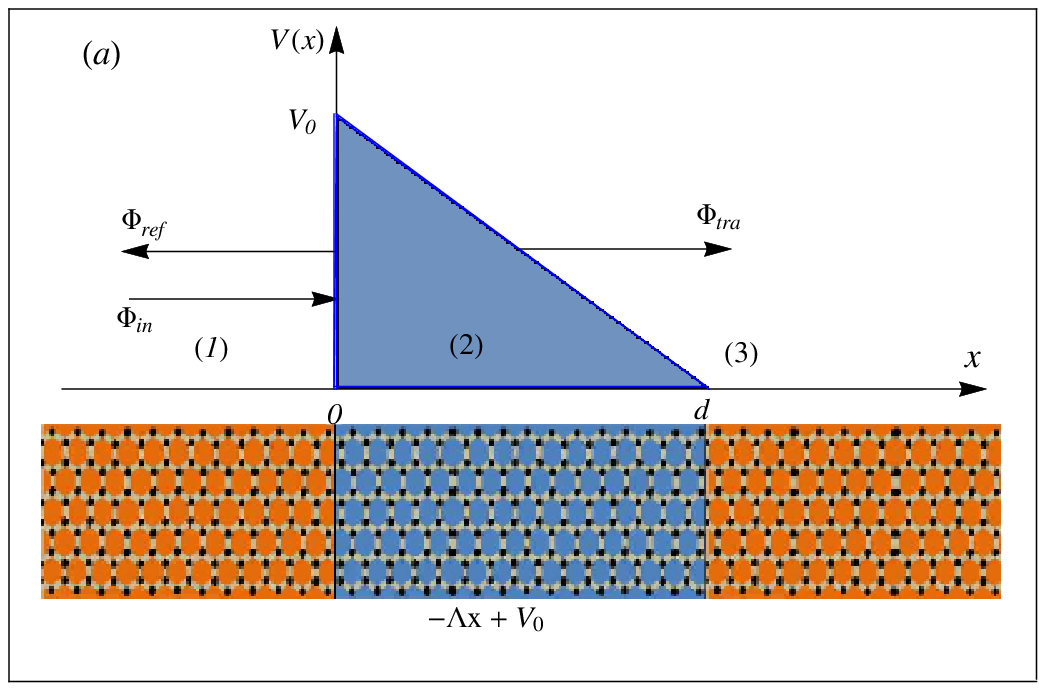}\ \ \ \ \ \
\includegraphics[width=8cm, height=5.5cm]{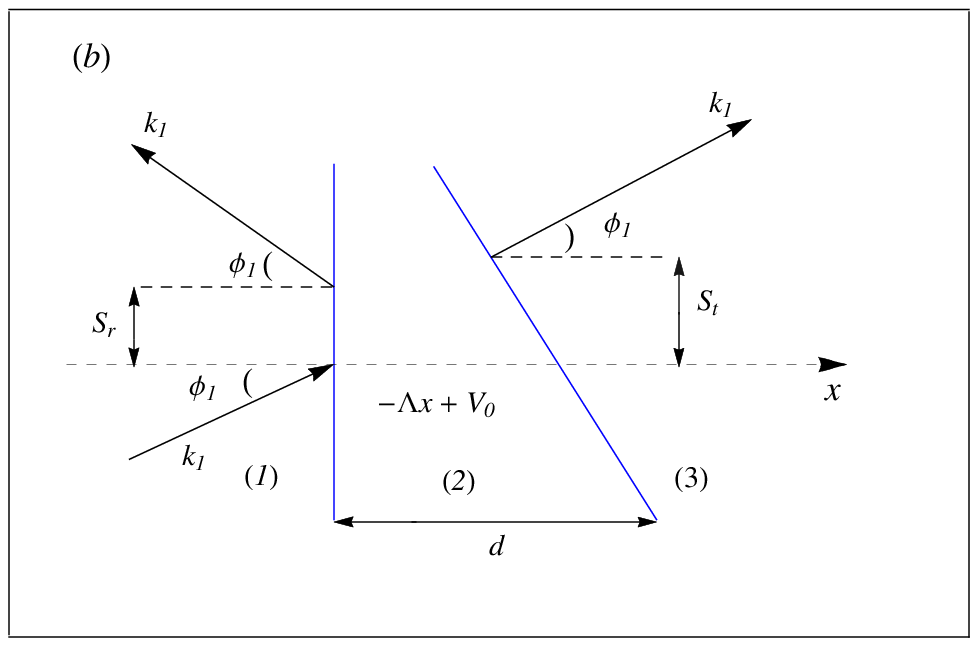}
 \caption{\sf{ (a): Profile of a linear barrier potential
 of width $d$ and height $V_0$ applied to graphene composed
 of three regions.
(b): The corresponding 
Goos-H\"anchan shifts in transmission and reflection.  
}}\lb{db.1}
\end{figure}

 The time-independent Dirac
equation for the spinor
$\Phi(x,y) $
at energy
$E=v_{F}\epsilon$ is given by
\begin{equation} \lb{eqh1}
\left[{\boldsymbol{\sigma}}\cdot\textbf{p}+v_j{\mathbb
I}_{2}\right]\Phi(x,y)=\epsilon \Phi(x,y)
\end{equation}
where we have rescaled the parameters $V_{j}=v_{F}v_{j}$, $\Lambda=v_{F}\varrho$, $\hbar=1$,
$V_{0}=v_{F}v_{0}$ as well as the linear  potential
\begin{equation}
v_j=
\left\{%
\begin{array}{ll}
    -\varrho x+v_0, & \hbox{$0\leq x\leq d$} \\
    0, & \hbox{otherwise} \\
\end{array}%
\right.
\end{equation}
 Our system is supposed to have finite width $W$
with infinite mass boundary conditions on the wavefunction at the
boundaries $y = 0$ and $y = W$ along the $y$-direction
\cite{Tworzydlo, Berry}. This results in a
quantization of the transverse momentum 
\begin{equation}
k_{y}=\frac{\pi}{W}\left(n+\frac{1}{2}\right),\qquad n=0,1,2 \cdots.
\end{equation}
Because of separability, we take the spinor
$\Phi_{j}(x,y)=\left(\varphi_{j}^{+}(x),\varphi_{j}^{-}(x)\right)^{\dagger}e^{ik_{y}y}$
and  solve the eigenvalue
equation to obtain the upper and lower components of the
eigenspinor in the incident and reflected region {1} ($x <
0$)
\begin{eqnarray}\label{eq3}
    && \Phi_{1} (x,y)=  \left(
            \begin{array}{c}
              {1} \\
              {z_{1}} \\
            \end{array}
          \right) e^{i(k_{1}x+k_{y}y)} + r\left(
            \begin{array}{c}
              {1} \\
               {-z_{1}^{-1}} \\
            \end{array}
          \right) e^{i(-k_{1}x+k_{y}y)}\\
        &&   z_{1} =s_{1}\frac{k_{1}+ik_{y}}{\sqrt{k_{1}^{2}+k_{y}^{2}}}\lb{eq7}
\end{eqnarray}
 where we have defined $s_{j}={\mbox{sign}}{\left(E\right)}$, and $r$ is a constant parameter
 while the corresponding dispersion relation can easily be obtained as
 \beq
\epsilon=s_1\sqrt{k_1^2 +k_y^2}.
\eeq
In region {2} ($0<x<d$),
we can express the general solution in terms of the parabolic
cylinder function \cite{HBahlouli} and then the first component is 
\begin{equation}\lb{hii1}
 \chi^{+}=c_{n1}
 D_{\nu_n-1}\left(Q\right)+c_{n2}
 D_{-\nu_n}\left(-Q^{*}\right)
\end{equation}
where $\nu_n=\frac{ik_{y}^{2}}{2\varrho}$,
$\epsilon_{0}=\epsilon-v_{0}$  and $
Q(x)=\sqrt{\frac{2}{\varrho}}e^{i\pi/4}\left(-\varrho
x+\epsilon_{0}\right) $, $c_{n1}$ and $c_{n2}$ are constants. We show that the second
one takes the form
\begin{eqnarray}\lb{hii2}
\chi^{-}=-\frac{c_{n2}}{k_{y}}\left[ 2(\epsilon_{0}- \varrho x)
 D_{-\nu}\left(-Q^{*}\right)
+
 \sqrt{2\varrho}e^{i\pi/4}D_{-\nu_n+1}\left(-Q^{*}\right)\right]
 -\frac{c_{n1}}{k_{y}}\sqrt{2\varrho}e^{-i\pi/4}
 D_{\nu_n-1}\left(Q\right).
\end{eqnarray}
The components of the spinor solution of the Dirac equation
\eqref{eqh1} in region { 2} can be derived from \eqref{hii1}
and \eqref{hii2} by setting
\beq
\varphi^{+}(x)=\chi^{+}+i\chi^{-}, \qquad \varphi^{-}(x)=\chi^{+}-i\chi^{-}
\eeq
which  give the spinor
\begin{eqnarray}
 \Phi_{2 } (x,y) &=& a_{1}\left(%
\begin{array}{c}
 \eta^{+}(x) \\
  \eta^{-}(x) \\
\end{array}%
\right)e^{ik_{y}y}+a_{2}\left(%
\begin{array}{c}
 \xi^{+}(x) \\
 \xi^{-}(x)\\
\end{array}%
\right)e^{ik_{y}y}
\end{eqnarray}
 where the functions $ \eta^{\pm}(x)$ and $\xi^{\pm}(x)$
read as
\begin{eqnarray}
\eta^{\pm}(x)&=&
 D_{\nu_{n}-1}\left(Q\right)\mp
 \frac{1}{k_{y}}\sqrt{2\varrho}e^{i\pi/4}D_{\nu_{n}}\left(Q\right)\\
\xi^{\pm}(x)&=&
 \pm\frac{1}{k_{y}}\sqrt{2\varrho}e^{-i\pi/4}D_{-\nu_{n}+1}\left(-Q^{*}\right)\nonumber\\
 &&
  \pm
 \frac{1}{k_{y}}\left(-2i\epsilon_{0}\pm
 k_{y}+2i \varrho x\right)D_{-\nu_{n}}\left(-Q^{*}\right)
\end{eqnarray}
$a_1$ and $a_2$ are two constant parameters.
Now solving the eigenvalue equation for the Hamiltonian
\eqref{eqh1} describing region {3} ($x
> d$), to end up with the spinor in the transmitted region
\begin{equation}\label{eq6}
 \Phi_{3} (x,y)= t \left(
            \begin{array}{c}
              {1} \\
              {z_{1}} \\
            \end{array}
          \right) e^{i(k_{1}x+k_{y}y)}
\end{equation}
where $t$ is a constant parameter and $z_1$ is given by \eqref{eq7}.

\section{Goos-H\"anshen shifts}

Before deriving the Goos-H\"anchen (GH) shifts in the beginning we  determine
the transmission and reflection probabilities
for electrons in graphene subject to a linear barrier potential.
For this,
we write the obtained eigenspinors in matrix notation and
 impose the continuity of the wavefunctions at each
interface of the linear barrier. We will use the above solutions to compute the corresponding
transmission and reflection coefficient $(t,r)$ associated to phase
shifts and build a bridge between quantum optics and Dirac fermions
in graphene. The continuity of the spinor
wavefunctions at each junction interface read as
\bqr \label{eq11}
\Phi_{ 1}(0)= \Phi_{2}(0), \qquad
\Phi_{2}(d)= \Phi_{3}(d).
\eqr
It is convenient to
express these relationships in terms of $2\times 2$
transfer matrices between different regions $M_{jj+1}$, which are 
\beq
\left(%
\begin{array}{c}
  a_{j} \\
  b_{j} \\
\end{array}%
\right)=M_{j, j+1}\left(%
\begin{array}{c}
  a_{j+1} \\
  b_{j+1} \\
\end{array}%
\right).
\eeq
After some algebra, we obtain the total transfer matrix
\beq
M=\prod_{j=1}^{4}M_{j j+1}
  \eeq
  as well as the relation between reflection and transmission amplitudes
  \begin{equation}
\label{systm1}
\left(%
\begin{array}{c}
  1 \\
  r \\
\end{array}%
\right)=M\left(%
\begin{array}{c}
  t \\
  0 \\
\end{array}%
\right)
\end{equation}
where $M_{12}$, $M_{2 3}$ are
transfer matrices that couple the wavefunction in the $j$-th
region to the wavefunction in the $(j + 1)$-th region. Explicitly, we have
\begin{eqnarray}
&&M=\left(%
\begin{array}{cc}
  m_{11} & m_{12} \\
  m_{21} & m_{22} \\
\end{array}%
\right)
\\
&&
M_{12}=\left(%
\begin{array}{cc}
  1 &1 \\
  z_{1} & -z^{\ast}_{1}  \\
\end{array}%
\right)^{-1}\left(%
\begin{array}{cc}
\eta^{+}(0) &  \xi^{+}(0)\\
 \eta^{-}(0) & \xi^{-} (0)\\
\end{array}%
\right)
\\
&&
M_{23}=\left(%
\begin{array}{cc}
 \eta^{+}(d) &  \xi^{+}(d)\\
 \eta^{-}(d) & \xi^{-} (d)\\
\end{array}%
\right)^{-1}\left(%
\begin{array}{cc}
  e^{\textbf{\emph{i}}k_{1} d} & e^{-\textbf{\emph{i}}k_{1} d} \\
  z_{1} e^{\textbf{\emph{i}}k_{1} d}  & -z_{1}^{\ast} e^{-\textbf{\emph{i}}k_{1} d}  \\
\end{array}%
\right)
\end{eqnarray}
with shorthand notations
\begin{equation}
\eta^{\pm}(0)=\eta_{0}^{\pm},\qquad
 \eta^{\pm}(d)=\eta_{d}^{\pm},\qquad
 \xi^{\pm}(0)=\xi_{0}^{\pm},\qquad
 \xi^{\pm}(d)=\xi_{d}^{\pm}.
\end{equation}
Combining all to end up with the transmission and reflection amplitudes
\begin{equation}\label{eq 63}
 t=\frac{1}{m_{11}}, \qquad  r=\frac{m_{21}}{m_{11}}
\end{equation}
 and after some
lengthy algebra, we can show that they take the explicit forms
\begin{eqnarray}
 t&=&\frac{e^{-ik_{1}d}\left[1+z_{1}^{2}\right]\left[\xi^{+}_{d}\eta^{-}_{d}-\xi^{-}_{d}\eta^{+}_{d}\right]}
{\left[\xi^{+}_{0}+z_1\xi^{-}_{0}\right]\left[\eta^{-}_{d}-z_1\eta^{+}_{d}\right]-
\left[\eta^{+}_{0}+z_1\eta^{-}_{0}\right]\left[\xi^{-}_{d}-z_1\xi^{+}_{d}\right]}
\\\nonumber\\
r&=&\frac{\eta^{+}_{0}\xi^{-}_{d}+z_1\left(\eta^{-}_{d}\xi^{-}_{0}+\eta^{+}_{0}\xi^{+}_{d}-\eta^{+}_{d}\xi^{-}_{0}
+\eta^{+}_{0}\xi^{-}_{d}\right)-z_{1}^{2}\left(\eta^{+}_{d}\xi^{+}_{0}-
\eta^{+}_{0}\xi^{+}_{d}-\eta^{-}_{d}\xi^{+}_{0}\right)}
{\left[\xi^{+}_{0}+z_1\xi^{-}_{0}\right]\left[\eta^{-}_{d}-z_1\eta^{+}_{d}\right]-
\left[\eta^{+}_{0}+z_1\eta^{-}_{0}\right]\left[\xi^{-}_{d}-z_1\xi^{+}_{d}\right]}
\end{eqnarray}
which be can written
 in complex notation as
\begin{equation}\lb{trref}
 t = \rho_t e^{i\varphi_{t}},
\qquad r = \rho_r e^{i\varphi_{r}}
\end{equation}
where the corresponding  phase shifts 
are given by
\begin{equation}
 \varphi_{t} = \arctan\left(i\frac{t^{\ast}-t}{t+t^{\ast}}\right),
 \qquad \varphi_{r} =
 \arctan\left(i\frac{(r^{\ast}-r)}{r+r^{\ast}}\right)
 \end{equation}
 as well as their modulus
 \begin{equation}
     \rho_{t} =\sqrt{\Re^2 [t]+\Im^2 [t]},
     \qquad \rho_{r} =\sqrt{\Re^2 [r])+\Im^2 [r]}.
 \end{equation}
Now from the above results, we can easily derive the corresponding
 transmission $T$ and reflection
$R$ probabilities as
\begin{equation}
  T=\rho_{t}^{2},\qquad  R=\rho_{r}^{2}.
\end{equation}


To study GH shifts in graphene scattered by linear barrier,
we consider
an incident, reflected and transmitted
beams around some transverse wavevector $k_y = k_{y_0}$ and
incident angle $\phi_{1}(k_{y_{0}})\in [0, \frac{\pi}{2}]$,
denoted by the subscript $0$. The incident one is given by
\begin{eqnarray}
   \Psi_{in}(x,y) &=& \int_{-\infty}^{+\infty}dk_y\ f(k_y-k_{y_0})\ e^{i(k_{1}(k_y)x+k_yy)}\left(
            \begin{array}{c}
              {1} \\
              {e^{i\phi_{1}(k_{y})}}
            \end{array}
          \right)\label{eq 79}
\end{eqnarray}
and the reflected takes the form
\begin{eqnarray}
\Psi_{re}(x,y) &=& \int_{-\infty}^{+\infty}dk_y\ r(k_y)\
f(k_y-k_{y_0})\ e^{i(-k_{1}(k_y)x+k_yy)}\left(
            \begin{array}{c}
              {1} \\
              {-e^{-i\phi_{1}(k_{y})}} \\
            \end{array}
          \right)\label{refl}
\end{eqnarray}
where the reflection amplitude is $r(k_y)=|r|e^{i\varphi_{r}}$ 
\eqref{trref}.
Here (\ref{eq 79}-\ref{refl})
are represented by writing the $x$-component of wavevector
$k_{1}$ and $\phi_{1}$ in terms  of the transverse wavevector $k_{y}$. 
The spinor plane waves are solutions of the eigenvalue equation \eqref{eqh1} and
$f(k_y-k_{y_0})$ is the angular spectral distribution, assumed of Gaussian shape
\begin{equation}
f(k_y-k_{y_0})=w_ye^{-w_{y}^2(k_y-k_{y_0})^2}
\end{equation}with $w_y$ is the half beam width at waist
\cite{Beenakker}. We can approximate the $k_{y}$-dependent terms
by a Taylor expansion around $k_{y}$  retaining only the first
order term to get
\begin{eqnarray}
&&\phi_{1}(k_{y})\approx
\phi_{1}(k_{y_{0}})+\frac{\partial\phi_{1}}{\partial
k_{y}}\Big|_{k_{y_{0}}}(k_{y}-k_{y_{0}})
\\
&&
k_{1}(k_{y})\approx k_{1}(k_{y_{0}})+\frac{\partial
k_{1}}{\partial k_{y}}\Big|_{k_{y_{0}}}(k_{y}-k_{y_{0}}).
\end{eqnarray}
In the same way, we write the transmitted beam as 
\begin{eqnarray}
\Psi_{tr}(x,y) &=& \int_{-\infty}^{+\infty}dk_y\ t(k_y)\
f(k_y-k_{y_0})\ e^{i(k_{1}(k_y)x+k_yy)}\left(
            \begin{array}{c}
              {1} \\
              {e^{i\phi_{1}(k_{y})}} \\
            \end{array}
          \right)\label{trans}
\end{eqnarray}
 where the transmitted coefficient
$t(k_y)=|t|e^{i\varphi_{t}}$ \eqref{trref} will be calculated
through the use of boundary conditions. The stationary-phase
approximation indicate that the GH shifts are equal to the
negative gradients of transmission and reflection phases with respect to $k_y$.
They are given by
 \begin{equation}
        S_{t}=- \frac{\partial \varphi_{t}}{\partial
        k_{y}}\Big|_{k_{y0}}, \qquad S_{r}=- \frac{\partial \varphi_{r}}{\partial
        k_{y}}\Big|_{k_{y0}}.
 \end{equation}

 Next we will numerically analyze and discuss the GH shifts for Dirac fermions in graphene
 scattered by a linear barrier.
 This will be done by tuning on different physical parameters characterizing
 our system under suitable conditions.

\section{Results and discussions}

Figure \ref{figm2}(a) shows the phase shifts $\varphi_{t}$ of the transmitted beam
versus energy potential $v_0$ for a linear
 barrier with incident energy $\epsilon=20$, $k_y=1$ and two different values of
 distance $d=5$ (red line), $d=10$ (green
  line). They are oscillating periodically from positive to negative values
  with the same amplitudes and change the phase as long as distance $d$
  increases.
Figure \ref{figm2}(b) presents
transmission and reflection  probabilities
as function of energy
potential
 $v_0$ with the same conditions as before.
We observe that $T$ takes maximum value for $v_0$ far away from incident energy
$\epsilon$. 
It then
decreases sharply for $v_{0}>\epsilon-2k_{y}$ until it reaches a
relative minimum and then begins to increase in an oscillatory
manner. The inverse behavior of $T$ is exhibited by the reflection $R$  showing
that the conservation  $T+R=1$ is well satisfied.

\begin{figure}[!ht]
\centering
\includegraphics[width=8cm, height=5.5cm]{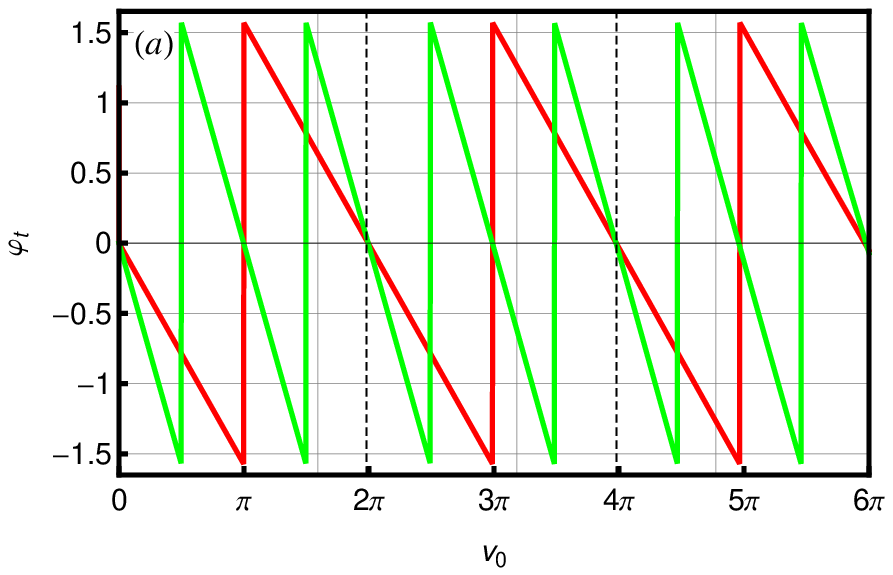}\ \ \ \
\includegraphics[width=8cm, height=5.5cm]{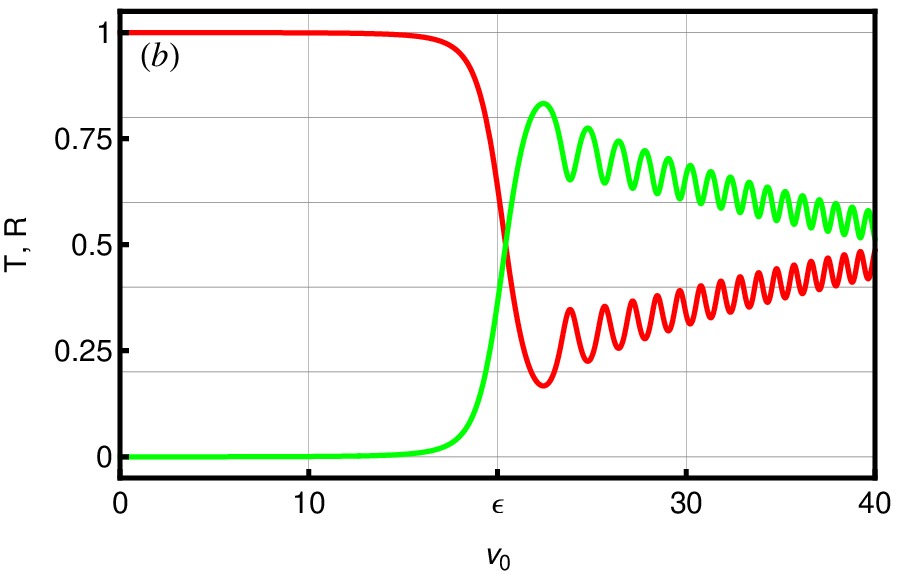}
 \caption{\sf{(Color online) (a): Phase shifts  $\varphi_{t}$ of the transmitted beam
 versus energy potential $v_{0}$  graphene linear barrier with
  $k_{y}=1$, $\epsilon=20$, $d=5$ (red line), $d=10$ (green
  line).
  (b): Transmission (red line) and reflection (green line) probabilities versus
 $v_{0}$ with
 $d=10$, $k_{y}=1$, $\epsilon=20$.}}\lb{figm2}
\end{figure}

 {Figure \ref{figm3} presents  the GH shifts $\{S_t,
S_r\}$ and probability $\{T, R\}$ as function of energy potential
$v_0$, with the parameters  $k_{y}=1$, $\epsilon=20$,
$d=0.2$ (magenta line), $d=2$ (blue line), $d=6$ (green line),
$d=10$ (red line).
From Figure
\ref{figm3}(a),
we observe
that $S_t$ increase to reach a peak as long as $v_0$ increases
and then rapidly decrease to zero after that it oscillate by
decreasing in negative regime.
In other words,
the GH shifts can be changed from positive to negative by
controlling the strength of the linear potential. However, the GH
shifts finally become positive with increasing the strength of the
energy potential $v_0$ up to $v_{0}>\epsilon$ and after
 decrease
in an oscillatory manner toward negative regime. From Figure \ref{figm3}(b)  one sees that
$S_r$
at
$v_{0}<\epsilon$ firstly become negative and null asymptotically,
after increase up to $\epsilon=v_{0}$. However in the opposite case, namely 
$v_{0}>\epsilon$, $S_r$ decrease and finally become negative then begin
to increase in an oscillatory manner. We clearly see that the GH shifts behave according to
the variation of the corresponding transmission and reflections probabilities as shown in
Figures \ref{figm3}c and \ref{figm3}d,  respectively.

\begin{figure}[!ht]
\centering
\includegraphics[width=8cm, height=5.5cm]{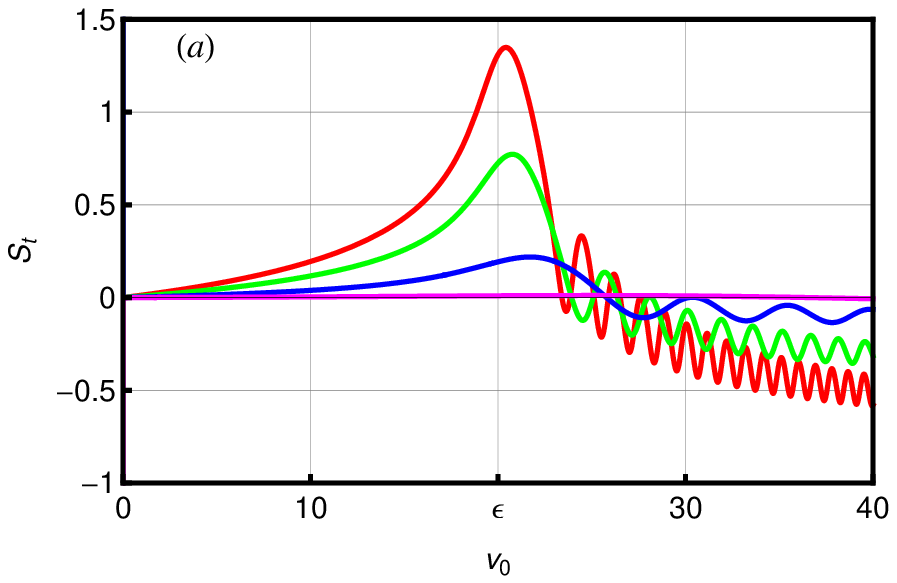}\ \ \ \
\includegraphics[width=8cm, height=5.5cm]{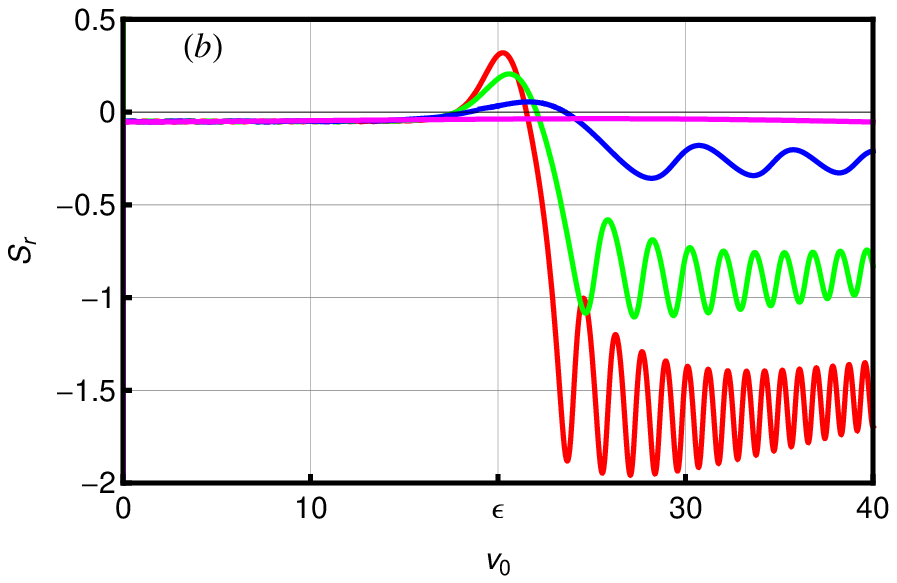}\\
\includegraphics[width=8cm, height=5.5cm]{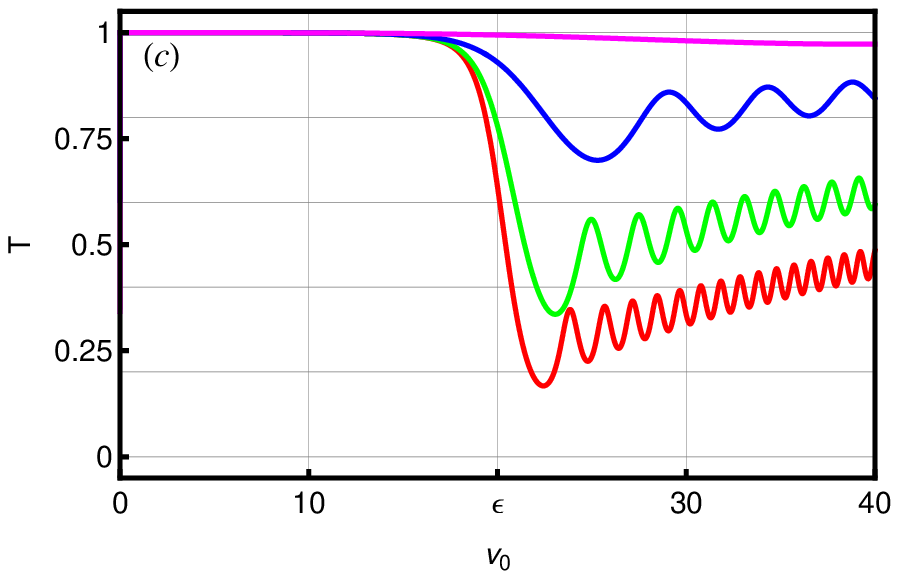}\ \ \ \
\includegraphics[width=8cm, height=5.5cm]{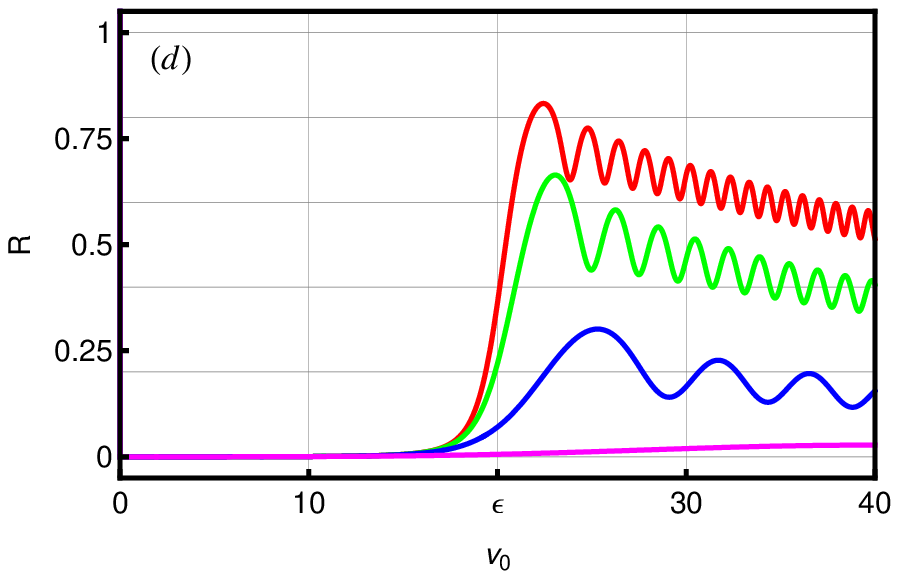}
 \caption{\sf{(color online) (a): GH shifts $S_{t}$ in transmission  and transmission
 probability $T$, (b): GH shifts $S_{r}$ in reflection  and reflection probability $R$ versus
energy potential $v_{0}$ with
   $k_{y}=1$, $\epsilon=20$, $d=0.2$ (magenta line), $d=2$ (blue line), $d=6$ (green line),
   $d=10$ (red line).}}\lb{figm3}
\end{figure}

\begin{figure}[!ht] \centering
\includegraphics[width=8cm, height=5.5cm]{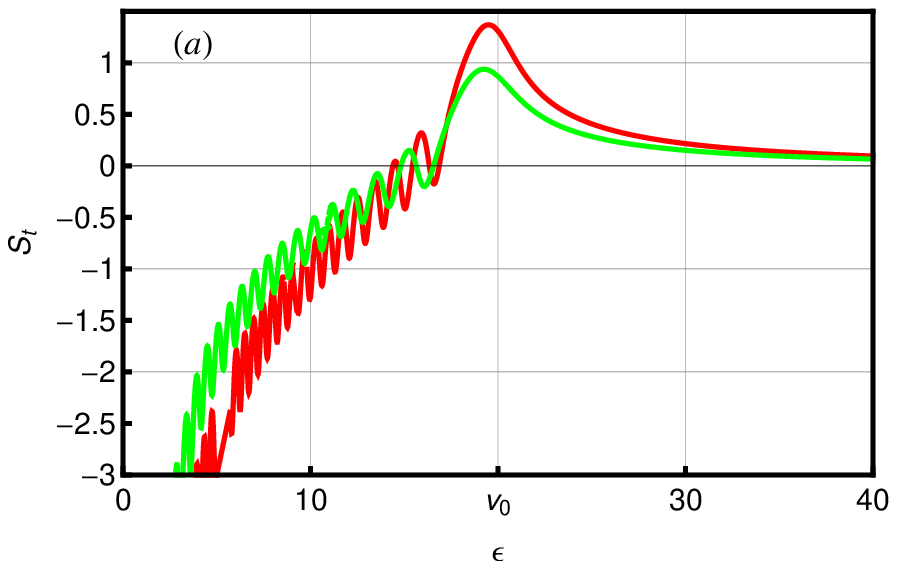}\ \ \ \
\includegraphics[width=8cm, height=5.5cm]{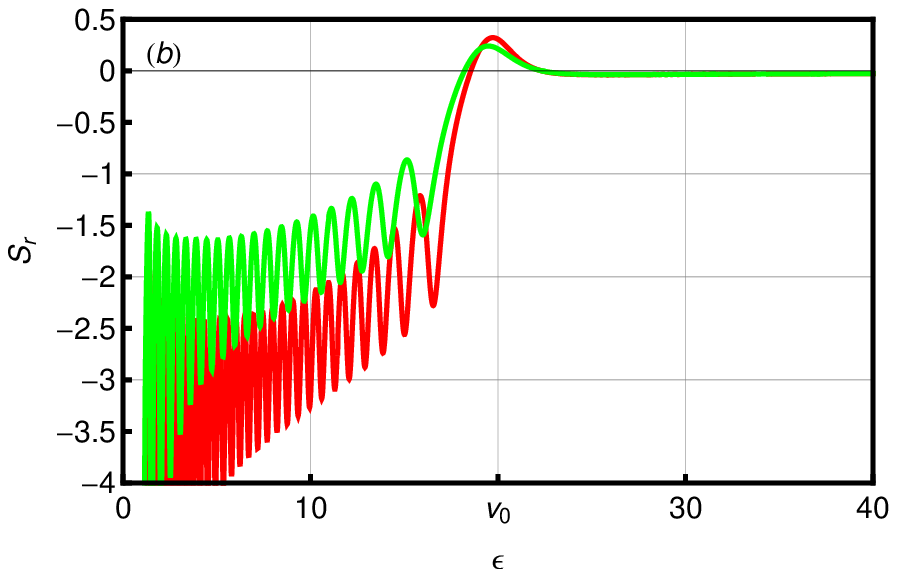}\\
\includegraphics[width=8cm, height=5.5cm]{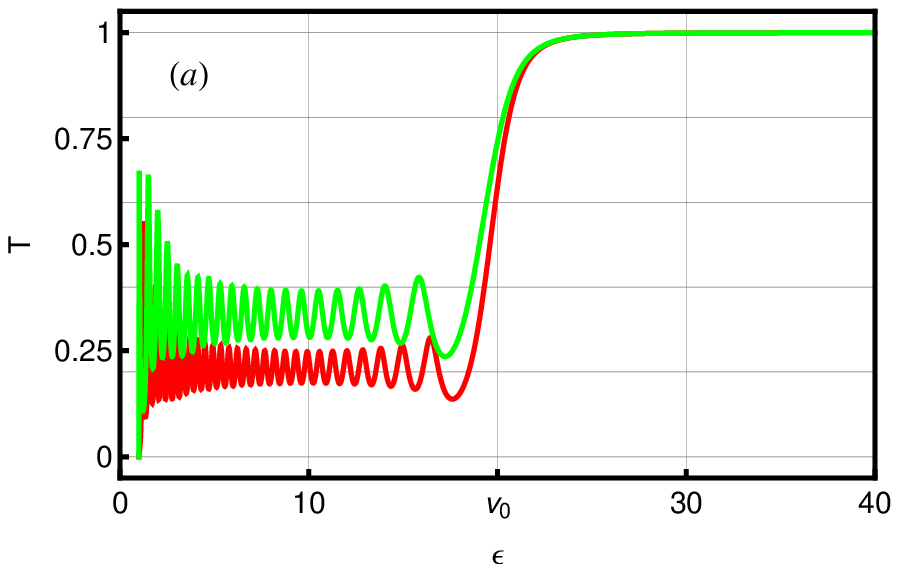}\ \ \ \
\includegraphics[width=8cm, height=5.5cm]{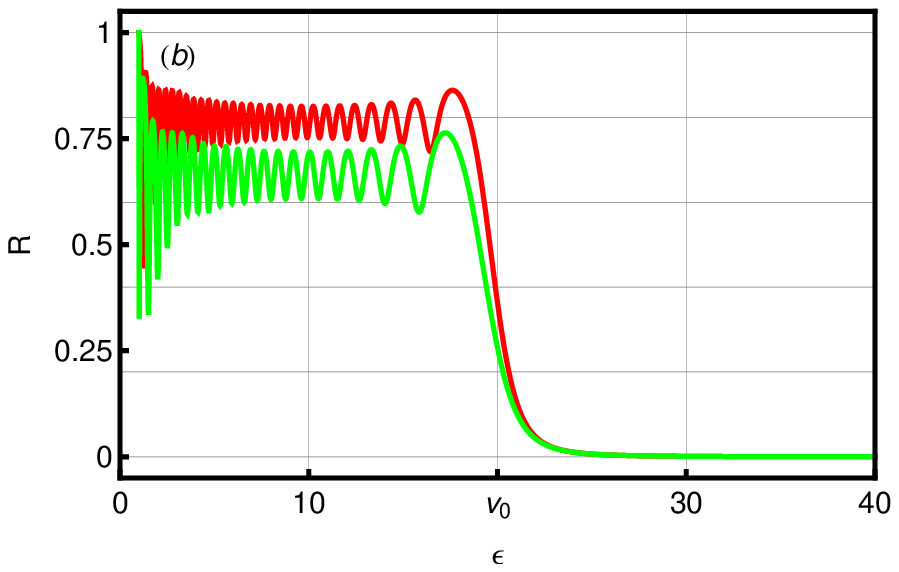}
 \caption{\sf{(color online) (a): GH shifts $S_{t}$ in transmission  and transmission
 probability $T$, (b): GH shifts $S_{r}$  in reflection
 and reflection probability $R$ versus incident energy $\epsilon$ with
 $k_{y}=1$, $v_{0}=20$, $d=7$ (green line),  $d=10$ (red line).}}\lb{figm4}
\end{figure}

 {In Figure \ref{figm4} we present  the GH shifts $\{S_t,
S_r\}$ and probability $\{T, R\}$ as function of energy
$\epsilon$, with  $k_{y}=1$, $v_0=20$, $d=7$
(green line), $d=10$ (red line). It is clearly seen that 
at the Dirac
points $\epsilon=v_0$, the GHL shifts change their signs. This
change shows clearly that they are strongly dependent on the
barrier heights.} In Figure \ref{figm4}(a),
 we observe that
the negative GH shifts $S_{t}$ in transmission at $\epsilon<v_{0}$ are negatively enhanced,
firstly increase accompanied with slight
oscillations and finally become positive with the strength of
 linear barrier. However, the positive GH shifts in
transmission at $\epsilon> v_{0}$ decrease, to reach asymptotically
a null value. We observe  that for a given energy,
$S_t$ in transmission
decrease if $d$ decreases and then
vanishes. {Note that below a certain critical energy
$\epsilon=k_y$ the transmission is almost zero, then it starts
oscillating  with frequency increases with increase of distance $d$, which is the size of the
region subject to the electric field.  Also the transmission
increases with $d$ decreases  in the range $k_y<\epsilon<v_0+2k_y$ and
reaches unity for energies above $\epsilon>v_0+2k_y$.}
 Figure \ref{figm4}(b) shows the GH shifts  $S_{r}$ in reflection and reflection probability
 $R$ as
function of incident energy  $\epsilon$. The negative and positive GH
shifts at $\epsilon<v_{0}$ are negatively enhanced and then begin
to increase in an oscillatory manner. However, the positive GH
shifts at $\epsilon> v_{0}$ firstly decrease and finally become
negative and null  asymptotically.  {We observe that  the reflection $R$
decreases a long as $d$ decreases and reaches zero value for energies above
$\epsilon>v_0+2k_y$.} We see that the GH shifts can be enhanced by
adjusting the incident energy or tuning on the distance of
region where the barrier is applied. This show that the GH shifts can be controlled
by linear potential.


\begin{figure}[!ht]
\centering
\includegraphics[width=8cm, height=5.5cm]{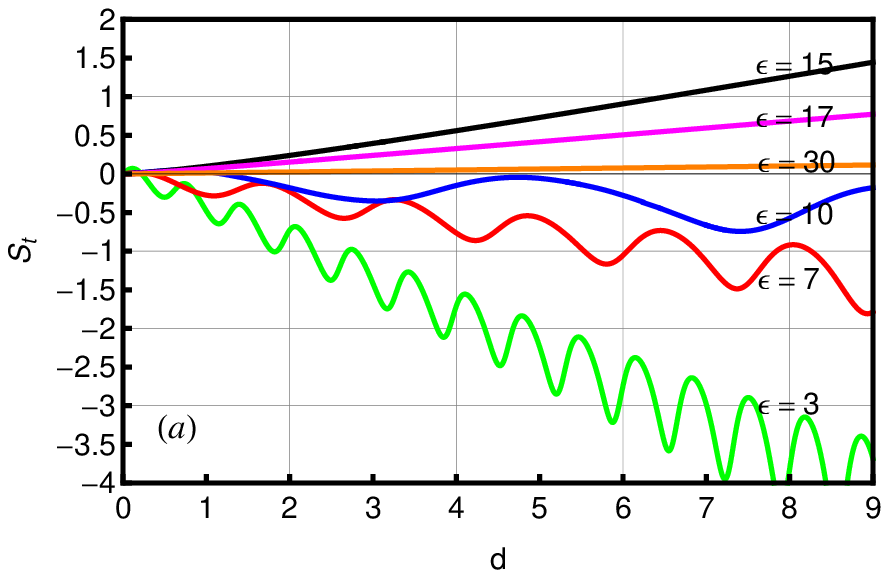}\ \ \ \
\includegraphics[width=8cm, height=5.5cm]{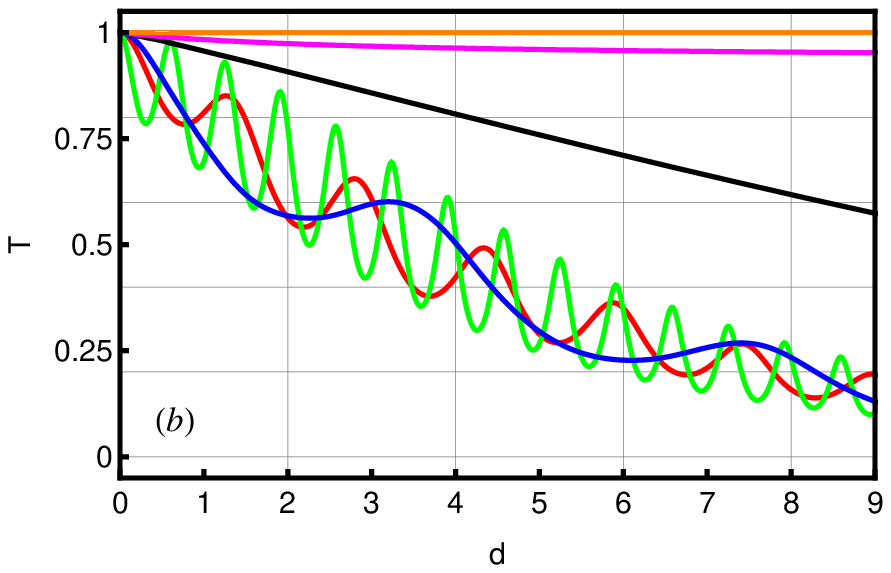}
 \caption{\sf{(color online) (a/b): GH shifts in transmission and transmission
 coefficient  $S_{t}/T$ versus barrier
width $d$, with
 $k_{y}=1$, $v_{0}=15$ and six values of incident energy $\epsilon=\{3, 7, 10, 15, 17, 30\}$.}}\lb{figm5}
\end{figure}

 Figure \ref{figm5} shows the GH shifts  $S_{t}$  in transmission and the corresponding transmission
 $T$ as function of  the barrier width $d$ with
 $k_{y}=1$, $v_{0}=15$ and different values of incident energy
 $\epsilon=\{3, 7, 10, 15, 17, 30\}$. In Figure \ref{figm5}(a), the
GH shifts  increase as long as $\epsilon$ increases in the range
$\epsilon<v_{0}$ with slight oscillations and finally become
negative.  However, in the range
$\epsilon\geq v_{0}$ the positive GH shifts 
 decrease once $\epsilon$ increases and there is
no oscillation. Figure \ref{figm5}(b) shows transmission
oscillating for $\epsilon<v_{0}$, which increases  with $\epsilon$ and is total
in the case $\epsilon\geq v_{0}$.

 {Figure \ref{figm6} presents the transmission,  GH shifts and phase
shifts as a function of the incident angle $\phi_1$ for
$\epsilon=2$, $v_0=5$ and different values of $d=\{3, 6, 10\}$.
 From Figure \ref{figm6}(a), 
we see that the perfect transmission occurs at
different angles and vice verse. It is observed that, the
transmission is always total for a normal incidence angle and
vanishes for specific values, it  decreases by increasing the
value of barrier width $d$. We observe that the curve of $T$ is
symmetric with respect to the normal incidence around the Dirac
point.
 From Figure \ref{figm6}(b) 
 one sees that
 the GH shifts can be changed from negative to positive
 according to variation of $\phi_1$. Indeed, the GH shifts are positive
 as long as the condition $\phi_1<0$ is
 satisfied and negative otherwise, namely $\phi_1>0$. The GH shifts are also symmetric with respect
to the normal incidence around the Dirac point.
In Figure \ref{figm6}(c), we observe that phase shifts 
are  oscillating according the value taken by transmission and show symmetric behavior.
All Figures show a symmetry at normal incidence angle $\phi_1=0$
separating positive and negative behavior of the GH shifts in transmission.

\begin{figure}[!ht]
\centering
\includegraphics[width=5.7cm, height=4cm]{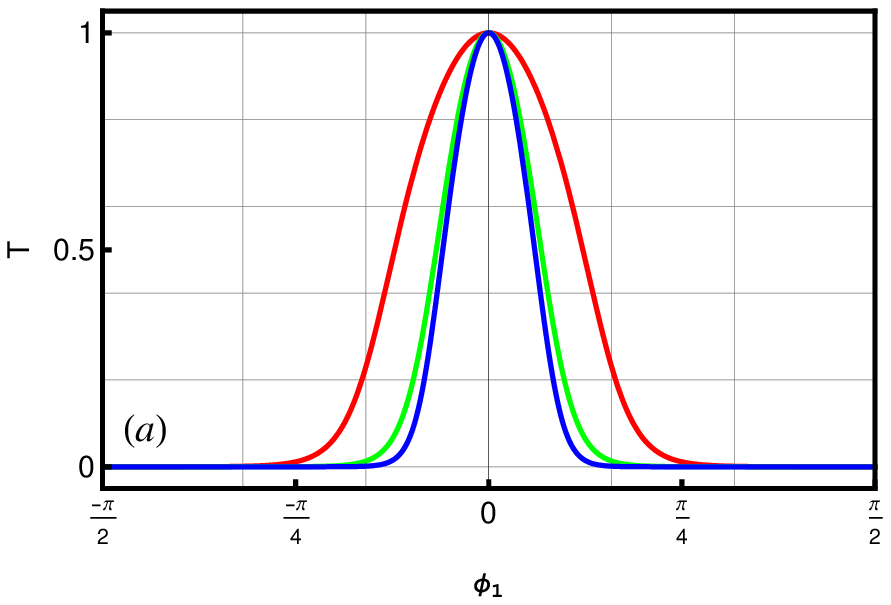}
\includegraphics[width=5.7cm, height=4cm]{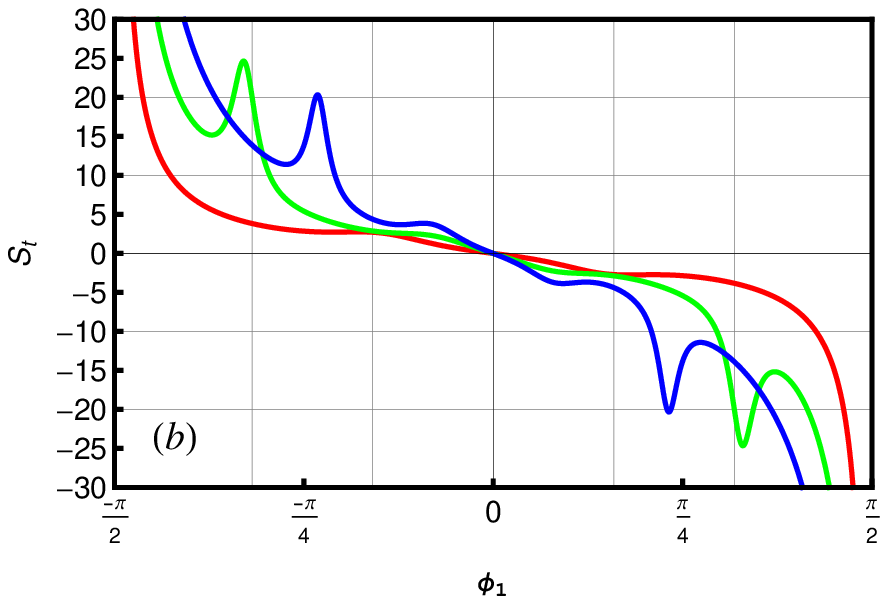}
\includegraphics[width=5.7cm, height=4cm]{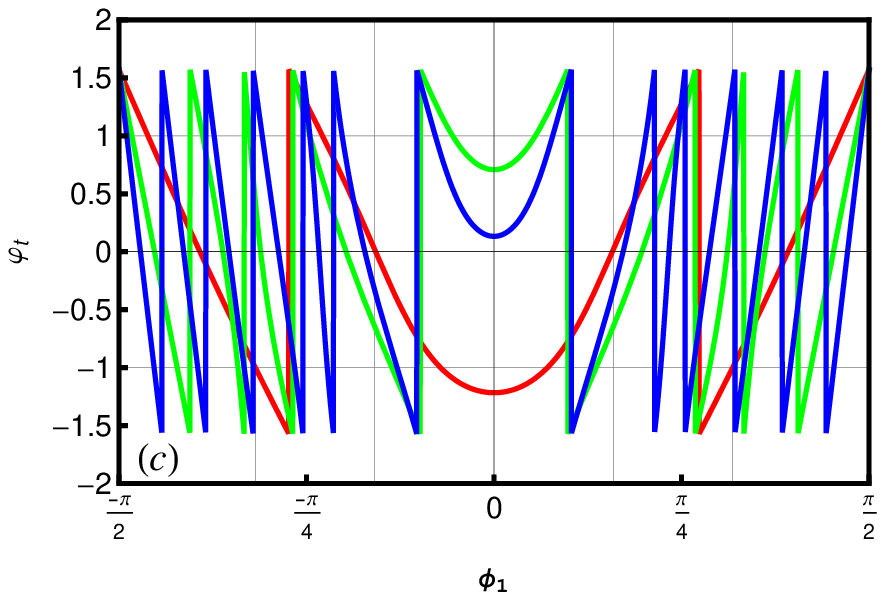}
 \caption{\sf{(color online) (a): Transmission probability $T(\phi_1)$, (b): GH shifts  in transmission
 $S_{t}(\phi_1)$, (c): phase shifts  $\varphi_{t}(\phi_1)$ versus
incidence angle $\phi_1$
with $v_{0}=5$, $\epsilon=2$, $d=3$ (red line),
$d=6$ (green line),  $d=10$ (blue line).}}\lb{figm6}
\end{figure}

\section{Conclusion}

We have studied the Goos-H\"anchen shifts for Dirac fermions in
graphene scattered by a linear barrier potential along the $x$-direction and  free
in the $y$-direction. After setting the solutions of the energy spectrum
for three regions composing our system,
we have
calculated the phase shifts of the transmitted and reflected beams
via the transmission and reflection coefficients.
These phases have been used
to  obtain the Goos-H\"anchen shifts in transmission and reflection
as function of a set of
physical parameters characterizing our
system.


Subsequently, we have presented different numerical results to
underline the basic features of our system. Indeed,
the phase shifts of transmitted beam showed oscillations with
the same amplitude,  as long as the barrier width increases
the phase shifts behave in the same way but with some  phases.
Later on
we have analyzed the GH shifts in terms of
incident energy, barrier width and energy potential. We have
observed that GH shifts are balancing from positive to negative
regimes by exhibiting different behaviors.
We hope that our graphene-like results could
present a possible way to simulate the quantum transport and GH
shifts 
of Dirac fermions in graphene through a linear barrier potential
by employing the optical technology.

\section*{Acknowledgment}

The generous support provided by the Saudi Center for Theoretical
Physics (SCTP) is highly appreciated by all authors.


\end{document}